\documentclass[pre,twocolumn]{revtex4-2}
\usepackage{amsmath}
\usepackage{amssymb}
\usepackage{graphicx}
\usepackage{braket}
\usepackage{stmaryrd}
\usepackage{hyperref}
\usepackage{color}
\usepackage{dsfont}

\hypersetup{
    colorlinks,
    citecolor=blue,
    linkcolor=blue,
    urlcolor=blue
}

\begin{document}
\title{Upper bound for entropy production in Markov processes}

\author{Tomohiro Nishiyama}
\email{htam0ybboh@gmail.com}
\affiliation{Independent Researcher, Tokyo 206-0003, Japan}

\author{Yoshihiko Hasegawa}
\email{hasegawa@biom.t.u-tokyo.ac.jp}
\affiliation{Department of Information and Communication Engineering, Graduate
School of Information Science and Technology, The University of Tokyo,
Tokyo 113-8656, Japan}
\date{\today}

\begin{abstract}
The second law of thermodynamics states that entropy production cannot be negative. Recent developments concerning uncertainty relations in stochastic thermodynamics, such as thermodynamic uncertainty relations and speed limits, have yielded refined second laws that provide lower bounds of entropy production by incorporating information from current statistics or distributions. 
In contrast, in this study, we bound the entropy production from above by terms comprising the dynamical activity and maximum transition-rate ratio. 
We derive two upper bounds: one applies to steady-state conditions, whereas the other applies to arbitrary time-dependent conditions. We verify these bounds through numerical simulation and identify several potential applications.
\end{abstract}

\maketitle
\section{Introduction\label{sec:Introduction}}

The second law of thermodynamics is considered one of the most fundamental and universal principles in physics and the cornerstone of many scientific disciplines.
It states that in any physical process, the entropy production of the system either increases or remains unchanged but never decreases.
The second law of thermodynamics was recently refined in stochastic thermodynamics through uncertainty relations, which have garnered attention in stochastic thermodynamics. 
One commonly cited example is the thermodynamic uncertainty relation (TUR) \cite{Barato:2015:UncRel,Gingrich:2016:TUP,Garrahan:2017:TUR,Dechant:2018:TUR,Terlizzi:2019:KUR,Hasegawa:2019:CRI,Hasegawa:2019:FTUR,Dechant:2020:FRIPNAS,Vo:2020:TURCSLPRE,Koyuk:2020:TUR,Pietzonka:2021:PendulumTURPRL,Brandner:2018:Transport,Carollo:2019:QuantumLDP,Guarnieri:2019:QTURPRR,Saryal:2019:TUR,Hasegawa:2020:QTURPRL,Sacchi:2021:BosonicTUR,Kalaee:2021:QTURPRE,Koyuk:2022:ManyBodyTUR,Hasegawa:2023:BulkBoundaryBoundNC}. The TUR states that achieving greater accuracy in thermodynamic systems comes at the cost of increased thermodynamic expenditures, such as entropy production or dynamical activity.
Furthermore, the TUR gives a bound for current fluctuations by the entropy production (or the dynamical activity).
From a different perspective, the TUR provides lower bounds of the entropy production, which are refinements of the second law of thermodynamics. 
Namely, the TUR states that
\begin{align}
    \Sigma\ge\frac{2\mathbb{E}[J]^{2}}{\mathrm{Var}[J]} \ge 0,
    \label{eq:Sigma_TUR}
\end{align}
where $\Sigma$ is the entropy production, $J$ is the thermodynamic current (see Section~\ref{sec:examples} for a detailed definition), and $\mathbb{E}[\cdot]$ and $\mathrm{Var}[J]$ denote the expectation value and variance, respectively. 
Equation~\eqref{eq:Sigma_TUR} provides a tighter bound to the conventional second law
by using additional information regarding the current. 
Another related uncertainty relation in stochastic thermodynamics is the classical speed limit (CSL)  \cite{Shiraishi:2018:SpeedLimit,Ito:2018:InfoGeo,Dechant:2019:Wasserstein,Ito:2020:TimeTURPRX,Nicholson:2020:TIUncRel,Vu:2021:GeomBound,Nakazato:2021:Wasserstein}. 
The CSL is a classical generalization of the quantum speed limit (QSL) \cite{Mandelstam:1945:QSL,Margolus:1998:QSL,Deffner:2017:QSLReview}, which places a limit on the speed of state change. 
An instance of the CSL is given by
\begin{align}
    \Sigma \ge k \mathcal{W}(P(\tau),P(0)),
    \label{eq:CSL}
\end{align}
where $P(0)$ and $P(\tau)$ are the initial and final densities, respectively, and 
$\mathcal{W}(\cdot,\cdot)$ is the Wasserstein distance between the probability densities at $t=0$ and $t=\tau$, and $k$ is a constant that does not depend on state. 
As with Eq.~\eqref{eq:Sigma_TUR}, the CLS
gives a lower bound of the entropy production and thus can be identified as an improved second law.

As demonstrated above, while the lower bound of entropy production has received considerable attention in uncertainty relations, its upper bound has been less explored.
The lower bound of the entropy production is of practical significance in thermodynamic inference because this method estimates entropy production based on trajectory measurements \cite{Seifert:2019:InferenceReview,Li:2019:EPInference,Manikandan:2019:InferEP}. 
However, the precision of these estimations remains relatively low for realistic cases. Reference~\cite{Roldan:2021:EPInference}, for example, estimated the entropy production based on a biological model; however, the numbers are often off by two or three digits. 
Therefore, if the upper bound is available, its inclusion can lead to more accurate estimates of the entropy production.
Moreover, once the upper bound of the entropy production can be derived,
several inequalities, such as Eqs. ~\eqref{eq:Sigma_TUR} and \eqref{eq:CSL} have an alternative upper bound other than the entropy production. 
The upper bound for the average entropy production was studied in Ref.~\cite{Limkumnerd:2017:EPUpperBound},
which is based on extrema of the entropy production \cite{Neri:2017:StopTimeST}. 
Reference~\cite{Salazar:2022:QEntUpperBound} derived an upper bound of quantum entropy production with the entropy flux rate.
Reference~\cite{di2023variance} derived a reversed TUR which gives an upper bound to the entropy production by the work exerted on a bead. 

In the present study, 
using results from Ref.~\cite{Vo:2022:UKTURJPA},
we show that entropy production
can be bounded from above by the terms comprising the dynamical activity [cf. Eq.~\eqref{eq:DA_A_def}], and the maximum ratio between any pair of transition rates [cf. Eq.~\eqref{eq:R_def}].
We derive two upper bounds, one that applies to steady-state conditions and another that applies to arbitrary time-dependent driven conditions. The latter bound appears to hold for general Markov processes with time-dependent transition rates, starting from any probability distribution.
By considering a long time limit with time-independent transition rates, it can been shown that the latter bound reduces to the former steady-state bound. 
We verify the obtained bounds using a numerical simulation. 
Moreover, we present possible applications of the obtained bounds. 

\section{Method and Results\label{sec:Results}}

\begin{figure}
\centering
\includegraphics[width=8.5cm]{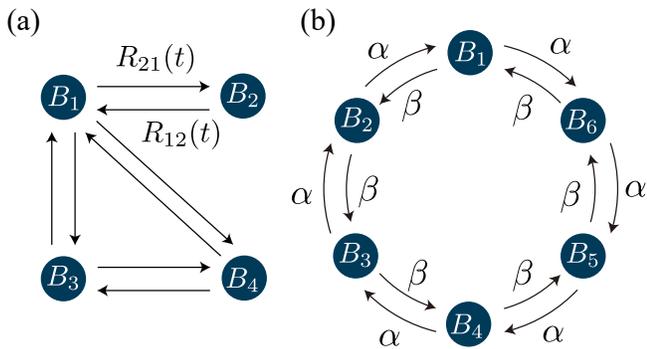}
\caption{
(a) Example of Markov processes considered in this study.
$R_{nm}(t)$ denote the transition rate from $B_m$ to $B_n$ at time $t$. 
If the transition from $B_m$ to $B_n$ is possible, then the transition in the opposite direction should also be possible
which is a requirement for defining entropy production. 
(b) Markov process with a ring topology that achieves the equality condition of Eq.~\eqref{eq:main_result_1}.
$\alpha>0$ and $\beta>0$ represent the transition rates for the clockwise and anti-clockwise directions, respectively. 
}
\label{fig:model}
\end{figure}

Consider a classical Markov process comprising $N$ states $\{B_1,B_2,\cdots,B_N\}$. 
Let $R_{nm}(t)$ be the transition rate from $B_m$ to $B_n$ at time $t$,
and let $P_m(t)$ be the probability of being $B_m$ at time $t$. 
The dynamics is supposed to obey the master equation:
\begin{align}
    \frac{d}{dt}P_{n}(t)=\sum_{m}R_{nm}(t)P_{m}(t),
    \label{eq:master_eq_def}
\end{align}
where we define the diagonal elements as $R_{nn}(t) \equiv -\sum_{m(\ne n)}R_{mn}(t)$.
We assume that if $R_{nm}(t) > 0$ for any indices $m$ and $n$ satisfying $m\ne n$, then $R_{mn}(t)>0$ [Fig.~\ref{fig:model}(a)].
Then, assuming local detailed balance, we can define the entropy production rate $\sigma(t)$ and dynamical activity rate $\mathfrak{a}(t)$ at time $t$:
\begin{align}
   \sigma(t)&\equiv\sum_{n<m}\left(P_{m}(t)R_{nm}(t)-P_{n}(t)R_{mn}(t)\right)\nonumber\\&\times\ln\frac{P_{m}(t)R_{nm}(t)}{P_{n}(t)R_{mn}(t)},\label{eq:EP_Sigma_def}\\\mathfrak{a}(t)&\equiv\sum_{n<m}\left(P_{m}(t)R_{nm}(t)+P_{n}(t)R_{mn}(t)\right).\label{eq:DA_A_def}
\end{align}
Here, the dynamical activity rate $\mathfrak{a}(t)$ quantifies the average number of jumps at time $t$. 
Moreover, we define the ratio $R$, which quantifies the maximum ratio between any pair of transition rates $R_{nm}(t)$:
\begin{align}
    R\equiv \max _{n \neq m, t} \frac{R_{m n}(t)}{R_{n m}(t)} \geq 1.
    \label{eq:R_def}
\end{align}
Note that, when $R_{nm} = R_{mn} = 0$, we define $R = 1$. 
When the system is time-independent, Eq.~\eqref{eq:R_def} reduces to 
\begin{align}
    R= \max _{n \neq m} \frac{R_{m n}}{R_{n m}} \geq 1.
    \label{eq:R_def2}
\end{align}

\subsection{Steady-state case}
We first consider a time-independent Markov process (i.e., $R_{nm}$ is time-independent) and assume that the system is in the steady state with $P_m^\mathrm{ss}$ being the steady-state distribution,
where $\sigma(t) = \sigma$ and $\mathfrak{a}(t)=\mathfrak{a}$. 
Under steady-state conditions, we obtain the following upper bound on the entropy production rate:
\begin{align}
    \sigma\leq\mathfrak{a}\left(\ln R\right)\frac{R-1}{R+1}.
    \label{eq:main_result_1}
\end{align}
Equation~\eqref{eq:main_result_1} is the first main result of this study and its proof is provided in Appendix~\ref{sec:derivation_ss}. 
Equation~\eqref{eq:main_result_1} provides an upper bound on the entropy production $\sigma$ in terms of the dynamical activity $\mathfrak{a}$ and the ratio $R$ [Eq.~\eqref{eq:R_def2}]. 
Let us consider the equality case of Eq.~\eqref{eq:main_result_1}.
Equation~\eqref{eq:main_result_1} becomes an equality when the entropy production and  dynamical activity rates are given by
\begin{align}
    \mathfrak{a}=c(\alpha+\beta),\quad\sigma=c(\beta-\alpha)\ln\frac{\beta}{\alpha},
    \label{eq:equality_condition}
\end{align}
for the transition rates $\alpha >0$ and $\beta>0$,
and  constant $c > 0$. 
Consider a Markov process with a ring topology, where the transition rates in the clockwise and anticlockwise directions are denoted by $\alpha$ and $\beta$, respectively [Fig.~\ref{fig:model}(b)]. 
Such a Markov process has been extensively studied in stochastic clock models \cite{Barato:2016:BrowClo}. 
The entropy production and dynamical activity of the ring system are given by Eq.~\eqref{eq:equality_condition} and hence Eq.~\eqref{eq:main_result_1} is saturated in the system. 

Equation~\eqref{eq:main_result_1} is meaningful from a practical point of view. 
Equation~\eqref{eq:main_result_1} shows that the ratio $\sigma/\mathfrak{a}$
can be bounded from above by $R$ alone. 
Since the measurement of $R$ is relatively easy, Eq.~\eqref{eq:main_result_1} can be used to infer the ratio $\sigma/\mathfrak{a}$ from measurements of stochastic trajectories. 

\subsection{Time-dependent driven case}

We have so far obtained the upper bound for entropy production for the steady-state case. 
Next, we derive the upper bound for entropy production in a time-dependent driven process. 
Suppose that the process begins at $t=0$ and ends at $t=\tau$ ($\tau > 0$). 
Let $\Sigma(\tau)$ and $\mathcal{A}(\tau)$ be the entropy production and dynamical activity, respectively, within the interval $[0,\tau]$:
\begin{align}
    \Sigma(\tau)&\equiv\int_{0}^{\tau}dt\,\sigma(t),\label{eq:Sigma_tau_def}\\
    \mathcal{A}(\tau)&\equiv\int_{0}^{\tau}dt\,\mathfrak{a}(t).\label{eq:A_tau_def}
\end{align}
$\Sigma(\tau)$ quantifies the  entropy production generated during the interval $[0,\tau]$, and
$\mathcal{A}(\tau)$ denotes the average number of jump events during $[0,\tau]$.
Then, we obtain upper bounds on the entropy production as follows:
\begin{align}
\Sigma(\tau)&\le \mathcal{A}(\tau)\left(\ln R\right)\frac{R-1}{R+1}+\left(1+\frac{2R\ln R}{R^2-1}\right)\ln N.
    \label{eq:main_result_2}
\end{align}
Equation~\eqref{eq:main_result_2} is the second result of this study,
which holds for an arbitrary Markov process starting from an arbitrary state.
Equation~\eqref{eq:main_result_2} shows that the upper bound of the entropy production is determined solely by $\mathcal{A}(\tau)$, $R$, and $N$. 
Note that \eqref{eq:main_result_2} trivially holds for $\Sigma(\tau)\le (1 + (\ln R) / (R+1) )\ln N$.
A proof for Eq.~\eqref{eq:main_result_2} is provided in Appendix~\ref{sec:derivation_driven}. 
The bound of Eq.~\eqref{eq:main_result_2} tightens when $\Sigma(\tau) \gg \ln N$.
Under this condition, Eq.~\eqref{eq:main_result_2} converges to the same inequality as derived for the steady-state case [Eq.~\eqref{eq:main_result_1}].
Therefore, in addition to $\Sigma(\tau) \gg \ln N$,
the conditions given by Eq.~\eqref{eq:equality_condition}
are required for a saturation of the bound of Eq.~\eqref{eq:main_result_2}]

\subsection{Numerical simulation}

\begin{figure}[ht]
\centering
\includegraphics[width=6cm]{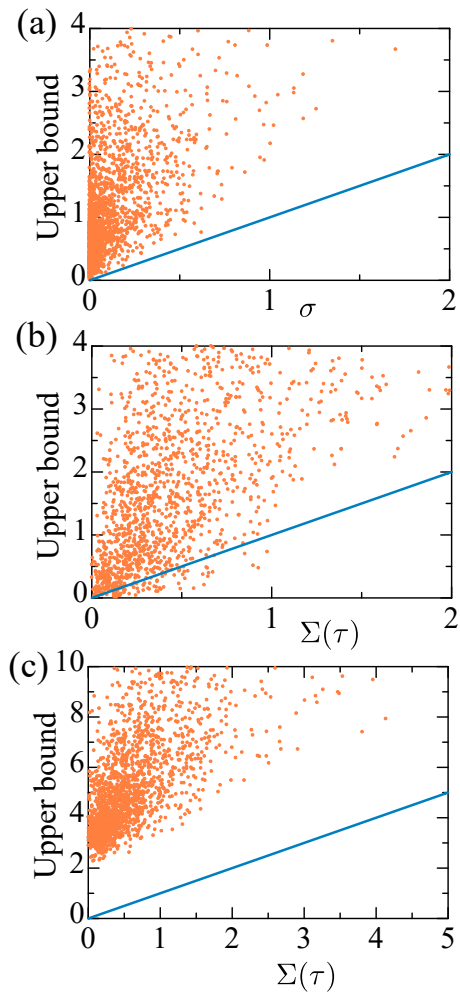}
\caption{
Numerical simulation of the upper bounds of entropy production.
(a) The right-hand side of Eq.~\eqref{eq:main_result_1} as a function of $\sigma$ under the steady-state condition. Random realizations are plotted by points and the solid line denotes the equality case. (b) The right-hand side of Eq.~\eqref{eq:main_result_1} as a function of $\sigma$ under non-steady-state conditions. 
Since Eq.~\eqref{eq:main_result_1} assumes the steady-state condition, the bound is not expected to hold under non steady-state conditions. 
The points and the solid line have the same meanings as those of (a).
(c) The right-hand side of Eq.~\eqref{eq:main_result_2} as a function $\Sigma(\tau)$ under non-steady-state conditions. 
The number of states is randomly sampled from $\{3,4,5\}$, and the transition rate $R_{nm}$ is randomly determined. 
In (a), the initial probability is $P_m^\mathrm{ss}$. 
In (b)--(c), the initial probability $P_m(0)$ is randomly sampled and the time duration $\tau$ is sampled from $[0.1, 5]$. 
}
\label{fig:random_case}
\end{figure}

Before providing applications, we verify the upper bounds [Eqs.~\eqref{eq:main_result_1} and \eqref{eq:main_result_2}] 
via numerical simulation. 
In this respect, we randomly generate Markov processes (time-independent transition rates $R_{nm}$) and calculate the entropy production rate $\sigma$ and its upper bounds. 
For details of the simulation parameters, please see the caption of Fig.~\ref{fig:random_case}. 
Figure~\ref{fig:random_case}(a) illustrates Eq.~\eqref{eq:main_result_1}, which is the upper bound for the steady-state condition by showing the entropy production rate $\sigma(\tau)$ as a function of the right-hand side of Eq.~\eqref{eq:main_result_1},
where the points represent random realizations and the solid line represents the saturating case. 
When Eq.~\eqref{eq:main_result_1} is satisfied, all points should be located above the solid line. As can be seen, all the points are above the line, indicating that Eq.~\eqref{eq:main_result_1} holds for the simulation. 
We also check whether Eq.~\eqref{eq:main_result_1} holds for out-of-steady state dynamics. 
In Fig.~\ref{fig:random_case}(b), we select random initial probability distributions and calculate the same quantities as in Fig.~\ref{fig:random_case}(a). 
We see that some points are below the solid line, implying that the steady-state bound [Eq.~\eqref{eq:main_result_1}] is not satisfied for the out-of-steady-state case. 

Next, we verify the bound for the out-of-steady-state condition [Eq.~\eqref{eq:main_result_2}]. 
Figure~\ref{fig:random_case}(c) visualizes Eq.~\eqref{eq:main_result_2},
where $\Sigma(\tau)$ is plotted as a function of the right-hand side of  Eq.~\eqref{eq:main_result_2}. 
In Fig.~\ref{fig:random_case}(c), the points denote random realizations, and the solid lines represent the saturating cases of Eq.~\eqref{eq:main_result_2}.  
Clearly, all points are located above the solid lines, numerically verifying the obtained bounds. 
Since Eq.~\eqref{eq:main_result_2} is trivially valid when $\Sigma(\tau)\le (1 + (\ln R) / (R+1) )\ln N$,
its tightness is weaker than the steady-state case [Eq.~\eqref{eq:main_result_1}],
which can be confirmed by comparing Figs.~\ref{fig:random_case}(a) and (c).

\section{Examples\label{sec:examples}}
In the first two examples, we show the new upper bounds for the precision of the generic current and the probability of entropy production.
\subsection{Precision of generic current}
\label{subsec:example_a}
We consider a system that is controlled by an arbitrary protocol $\lambda(vt)$ with speed parameter $v$, and let $R_{nm}(t)=R_{nm}(\lambda(vt))$ for all $m$ and $n$. Let $\omega_\tau =\{n_0, (n_1, t_1), \cdots, (n_K, t_K)\}$
be a stochastic trajectory of the system during the time interval $[0, \tau]$, where the system is initially in state $n_0$ and a transition from state $n_{i-1}$ to state $n_i$ occurs at time $t_i$ for each $1\le i\le K$. 
For each trajectory, we consider the generic current $J=J_d(\omega_\tau)\equiv\sum_{i=1}^K d_{n_i n_{i-1}}$for anti-symmetric coefficient $d_{mn}=-d_{nm}$, and we define the precision of the generic current as $\mathfrak{p}(J)\equiv (\nabla \mathbb{E}[J])^2 / \mathrm{Var}[J]$. 
Here, $\nabla\equiv \tau\partial_\tau - v\partial_v$ is a differential operator, and $\mathbb{E}[\cdot]$ and $\mathrm{Var}[\cdot]$ denote the ensemble average and variance of the current, respectively. 
Using the thermodynamic and kinetic uncertainty relation in Ref.~\cite{Vo:2022:UKTURJPA}, we obtain
\begin{align}
    \frac{\mathfrak{p}(J)}{\mathcal{A}(\tau)}\le\frac{\Sigma(\tau)^2}{4\mathcal{A}(\tau)^2{f\left(\frac{\Sigma(\tau)}{2\mathcal{A}(\tau)}\right)}^2},
\end{align}
where $f(x)$ is the inverse function of $x\tanh x$. 
Combining $g(x)=x/f(x)$, where $g:[0,\infty) \rightarrow [0,1)$ is the inverse function of $x\tanh^{-1}x$ in Ref.~\cite{nishiyama2022thermodynamic}, we obtain
\begin{align}
     \frac{\mathfrak{p}(J)}{\mathcal{A}(\tau)}\le g\left(\frac{\Sigma(\tau)}{2\mathcal{A}(\tau)}\right)^2.
\end{align}
Since $g$ increases monotonically, by combining [Eqs.~\eqref{eq:main_result_1} and \eqref{eq:main_result_2}], 
it follows that
\begin{align}
    \frac{\mathfrak{p}(J)}{\mathcal{A}(\tau)} \le g\left(\frac{\ln R}2\cdot \frac{R-1}{R+1} + \frac{C}{2\mathcal{A}(\tau)} \right)^2,
    \label{eq:upperbound_precision}
\end{align}
where the constant $C$ on the right-hand side is equal to zero for the steady-state case, and $C=\left(1+ 2R\ln R / (R^2-1)\right)\ln N$ 
for the time-dependent driven case.
Since $|g(x)|<1$ for all non-negative $x$, this inequality is tighter than the kinetic uncertainty relation $\mathfrak{p}(J) / \mathcal{A}(\tau)\le 1$ in ~\cite{Terlizzi:2019:KUR}. For the steady-state case, note that the right-hand side of Eq.~\eqref{eq:upperbound_precision} depends only on $R$, which is determined by the system. When $d_{mn}=1$ for $n<m$, since $\mathbb{E}[J]=c\tau(\beta-\alpha)$ and $\mathrm{Var}[J]=c\tau(\alpha+\beta)$ hold in the case presented in Fig.~\ref{fig:model}(b), Eq.~\eqref{eq:upperbound_precision} becomes the equality.

\subsection{Probability of entropy production}
We derive an upper bound for the probability that the entropy production is greater than or equal to a given value $s\geq 0$. We also derive an upper bound for the probability that the entropy production is less than or equal to $s$. We assume that the strong detailed fluctuation theorem holds, $P(\Sigma)/P(-\Sigma) = e^\Sigma$,
where $P(\Sigma)$ is the probability of entropy production $\Sigma$. 
To satisfy this condition, the system must meet two requirements \cite{Spinney:2013:FTReview}. First, the initial and final probability distributions must agree. Second, the external protocol must be time symmetric. 
Since $\phi(\Sigma)\equiv \Sigma(1-\exp(-\Sigma))$ is non-decreasing and non-negative for $\Sigma\geq 0$, we obtain
 \begin{align}
     \phi(s)P(\Sigma\geq s)&\le \int_s^\infty d\Sigma \phi(\Sigma) P(\Sigma) \nonumber\\\
     &\le \int_0^\infty d\Sigma \phi(\Sigma) P(\Sigma)\nonumber\\&=\int_{-\infty}^\infty d\Sigma \Sigma P(\Sigma)=\Sigma(\tau),
 \end{align}
 where we use $P(-\Sigma)=\exp(-\Sigma)P(\Sigma)$. 
 By combining this relation with the upper bounds [Eqs.~\eqref{eq:main_result_1} and \eqref{eq:main_result_2}],
we obtain
 \begin{align}
     P(\Sigma\geq s)\le\frac{\mathcal{A}(\tau)\ln R\cdot\frac{R-1}{R+1}+C}{s(1-\exp(-s))},
    \label{eq:upperbound_probability}
 \end{align}
where the definition of $C$ is the same as in Example~\ref{subsec:example_a}.
From Eq.~\eqref{eq:upperbound_probability}, we obtain 
  \begin{align}
    P(\Sigma\le-s)&=\int_{s}^{\infty}d\Sigma\exp(-\Sigma)P(\Sigma)\nonumber\\&\le\exp(-s)P(\Sigma\geq s)\nonumber\\&\le\frac{\mathcal{A}(\tau)\ln R\cdot\frac{R-1}{R+1}+C}{s(\exp(s)-1)}.
    \label{eq:Sigma_s_bound}
 \end{align}

\subsection{Arrow of time inference}
As an example of the entropy-production upper bound, we consider the inference of the arrow of time from trajectory measurements 
\cite{Jarzynski:2011:IrrIneq,Seif:2021:MachineLearning}.
Suppose that you are shown two movies. The first is a forward movie depicting a system that undergoes a process in which the external protocol $\lambda$ changes from $A$ to $B$. The other is a backward movie that displays the reverse process, where $\lambda$ goes from $B$ to $A$, and is being played backward.
The initial distribution of the backward process is identical to the final probability of the forward process, as assumed in the fluctuation theorem. 
Our task is to guess whether this movie is the forward or backward movie. 
It is known that the likelihood of the forward process given a trajectory $\omega_\tau$ is 
\begin{align}
    P(F|\omega_\tau)=\frac{1}{1+e^{-\Sigma}}.
    \label{eq:prob_F_Gamma}
\end{align}
Equation~\eqref{eq:prob_F_Gamma} demonstrates that the direction of time can be inferred more accurately by a larger entropy production.
Using Eq.~\eqref{eq:main_result_2}, we can obtain the upper bound of the likelihood:
\begin{align}
    P(F|\omega_\tau)\le\left(R^{-\frac{\mathcal{A}(\tau)\left(R-1\right)}{R+1}}N^{\frac{-R^2-2R\ln\left(R\right)+1}{R^2-1}}+1\right)^{-1},
    \label{eq:PFgamma_relation}
\end{align}
for  $\Sigma(\tau)> \left(1 + (\ln R) / (R+1) \right) \ln N$.

\section{Conclusion\label{sec:conclusion}}

In this study, we derived the upper bounds for entropy production based on the dynamical activity and the maximum transition-rate ratio. We established two upper bounds, one for steady-state conditions and another for arbitrary time-dependent conditions. Furthermore, we performed numerical simulations to confirm the validity of these limits. We also identified several potential applications of the upper bounds.
We expect that these findings will improve our understanding of nonequilibrium dynamics given that entropy production and dynamical activity are fundamental to thermodynamics.
Finally, we discuss the application of our upper bound to real experimental data. 
To accurately observe the upper bound of the entropy production derived from experimental data, we must address several challenges. In real-world scenarios, observations commonly capture only a portion of a Markov chain. Consequently, we cannot calculate $R$, which makes it impossible to implement the proposed method directly. 
Resolving this issue is a topic for future studies.

\appendix

\section{Derivation of steady-state case\label{sec:derivation_ss}}

We provide a derivation of Eqs.~\eqref{eq:main_result_1},
which holds true under steady-state conditions. 
The entropy production rate $\sigma$ [Eq.~\eqref{eq:EP_Sigma_def}] admits the following relation:
\begin{align}
    \sigma&=\sum_{n<m}\left(P_{m}^{\mathrm{ss}}R_{nm}-P_{n}^{\mathrm{ss}}R_{mn}\right)\ln\frac{P_{m}^{\mathrm{ss}}R_{nm}}{P_{n}^{\mathrm{ss}}R_{mn}}\nonumber\\&=\sum_{n<m}\left(P_{m}^{\mathrm{ss}}R_{nm}-P_{n}^{\mathrm{ss}}R_{mn}\right)\ln\frac{R_{nm}}{R_{mn}}\nonumber\\&\leq\left(\max_{m\ne n}\ln\frac{R_{nm}}{R_{mn}}\right)\sum_{n<m}\left|P_{m}^{\mathrm{ss}}R_{nm}-P_{n}^{\mathrm{ss}}R_{mn}\right|\nonumber\\&=\ln R\sum_{n<m}\left|P_{m}^{\mathrm{ss}}R_{nm}-P_{n}^{\mathrm{ss}}R_{mn}\right|,
    \label{eq:sigma_UB}
\end{align}
where $R$ is defined in Eq.~\eqref{eq:R_def}.
Using the inequality derived in Ref.~\cite{Vo:2022:UKTURJPA}, we obtain
\begin{align}
  \frac{\sigma}{f\left(\frac{\sigma}{2\mathfrak{a}}\right)}\geq2\sum_{n<m}\left|P_{m}^{\mathrm{ss}}R_{nm}-P_{n}^{\mathrm{ss}}R_{mn}\right|,
  \label{eq:sigma_f_eq}
\end{align}
where $f(x)$ is the inverse function of $x \tanh x$ as defined in the main text. 
Substituting Eq.~\eqref{eq:sigma_f_eq} into Eq.~\eqref{eq:sigma_UB}, we obtain
\begin{align}
    \frac{\ln R}{2}\geq f\left(\frac{\sigma}{2\mathfrak{a}}\right).
\end{align}
Because $x \tanh x$ is a monotonically increasing function, we obtain Eq.~\eqref{eq:main_result_1}. 

\section{Derivation of time-dependent driven case\label{sec:derivation_driven}}

Here, we provide a derivation of Eq.~\eqref{eq:main_result_2}. 
The following derivation assumes that
\begin{align}
    \Sigma(\tau)>\left(1+\frac{\ln R}{R+1}\right)\ln N.
    \label{eq:Sigma_assumption}
\end{align}
As will be shown, the assumption of Eq.~\eqref{eq:Sigma_assumption} is automatically satisfied by the final inequality [Eq.~\eqref{eq:non_ss_final_ineq}]. 
From the definition of $\Sigma(\tau)$, the following relation holds:
\begin{align}
    \Sigma(\tau)=&\int_{0}^{\tau}dt\sum_{n<m}\left(P_{m}(t)R_{nm}(t)-P_{n}(t)R_{mn}(t)\right)\nonumber\\&\times\ln\frac{P_{m}(t)R_{nm}(t)}{P_{n}(t)R_{mn}(t)}\nonumber\\=&\int_{0}^{\tau}dt\sum_{n<m}\left(P_{m}(t)R_{nm}(t)-P_{n}(t)R_{mn}(t)\right)\ln\frac{R_{nm}(t)}{R_{mn}(t)}\nonumber\\&+S(\tau)-S(0)\nonumber\\\le&\left(\ln R\right)\int_{0}^{\tau}dt\sum_{n<m}\left|P_{m}(t)R_{nm}(t)-P_{n}(t)R_{mn}(t)\right|\nonumber\\&+\ln N,
    \label{eq:Sigma_tau_ineq}
\end{align}
where $S(t)$ is the Shannon entropy at time $t$, $S(t) \equiv -\sum_n P_n(t) \ln P_n(t)$. 
$S(\tau)-S(0)$ in the second line of Eq.~\eqref{eq:Sigma_tau_ineq} can be derived from
\begin{align}
    &S(\tau)-S(0)\nonumber\\&=\int_{0}^{\tau}\frac{d}{dt}S(t)dt\nonumber\\&=\int_{0}^{\tau}dt\sum_{n<m}\left(P_{m}(t)R_{nm}(t)-P_{n}(t)R_{mn}(t)\right)\ln\frac{P_{m}(t)}{P_{n}(t)}.
    \label{eq:sigma_f_supplement}
\end{align}
In the last inequality in Eq.~\eqref{eq:Sigma_tau_ineq}, we used the relation $0 \le S \le \ln N$.
From Eq.~(27) in Ref.~\cite{Vo:2022:UKTURJPA}, the following relation holds:
\begin{align}
    \frac{\Sigma(\tau)}{f\left(\frac{\Sigma(\tau)}{2\mathcal{A}(\tau)}\right)}\geq2\int_{0}^{\tau}dt\sum_{n<m}\left|P_{m}(t)R_{nm}(t)-P_{n}(t)R_{mn}(t)\right|.
    \label{eq:sigma_f_ineq2}
\end{align}
Substituting Eq.~\eqref{eq:sigma_f_ineq2} into Eq.~\eqref{eq:Sigma_tau_ineq}, we obtain
\begin{align}
    \Sigma(\tau)-\ln N\le\frac{\ln R}{2f\left(\frac{\Sigma(\tau)}{2\mathcal{A}(\tau)}\right)}\Sigma(\tau).
    \label{eq:sigma_f_ineq3}
\end{align}
From Eq.~\eqref{eq:Sigma_assumption}, the left-hand side of Eq.~\eqref{eq:sigma_f_ineq3} appears to be positive. 
Thus, we have
\begin{align}
    f\left(\frac{\Sigma(\tau)}{2\mathcal{A}(\tau)}\right)\leq\frac{\ln R}{2\left(\Sigma(\tau)-\ln N\right)}\Sigma(\tau)=\frac{\ln R}{2}(Y(\tau)+1),
    \label{eq:sigma_f_ineq4}
\end{align}
for $\Sigma(\tau) - \ln N > 0$, where
\begin{align}
    Y(\tau)\equiv\frac{\Sigma(\tau)}{\Sigma(\tau)-\ln N}-1=\frac{\ln N}{\Sigma(\tau)-\ln N}>0.
    \label{eq:Ytau_def}
\end{align}
By simply considering $f^{-1}(x)$ in Eq.~\eqref{eq:sigma_f_ineq4}, we obtain
\begin{align}
    \frac{\Sigma(\tau)}{2\mathcal{A}(\tau)}\leq\frac{\ln R}{2}(Y(\tau)+1)\frac{R-R^{-Y(\tau)}}{R+R^{-Y(\tau)}}.
    \label{eq:sigma_f_ineq5}
\end{align}
Since $R^{-Y(\tau)}=\exp (-Y(\tau) \ln R) \geq 1-Y(\tau) \ln R$, Eq.~\eqref{eq:sigma_f_ineq5} can be bounded as follows: 
\begin{align}
\frac{\Sigma(\tau)}{2\mathcal{A}(\tau)}&\le\frac{\ln R}{2}(Y(\tau)+1)\frac{R-R^{-Y(\tau)}}{R+R^{-Y(\tau)}}\nonumber\\&\leq\frac{\ln R}{2}(Y(\tau)+1)\frac{R-1+Y(\tau)\ln R}{R+1-Y(\tau)\ln R},
\label{eq:sigma_f_ineq6}
\end{align}
where we use $R+1-Y(\tau)\ln R>0$ based on the assumption of  Eq.~\eqref{eq:Sigma_assumption} and $R\ge 1$.
Substituting $\Sigma(\tau)=(\ln N)(Y(\tau)+1)/Y(\tau)$ into Eq.~\eqref{eq:sigma_f_ineq6}, we obtain
\begin{align}
    &(R+1-Y(\tau)\ln R)\ln N\nonumber\\&\leq\mathcal{A}(\tau)Y(\tau)\ln R\left(R-1+Y(\tau)\ln R\right).
    \label{eq:sigma_f_ineq7}
\end{align}
Substituting $X(\tau)=1/Y(\tau)$ into Eq.~\eqref{eq:sigma_f_ineq7}, we obtain the following quadratic equation with respect to $X(\tau)$:
\begin{align}
&(R+1)(\ln N) X(\tau)^{2}-\left(\mathcal{A}(\tau)(R-1) + \ln N \right)(\ln R)X(\tau)\nonumber\\
    &-\mathcal{A}(\tau)(\ln R)^{2}\equiv DX(\tau)^2-EX(\tau)-F\le 0,
    \label{eq:sigma_f_ineq8}
\end{align}
where $D\equiv (R+1)\ln N$, $E\equiv (\mathcal{A}(\tau)(R-1) + \ln N )\ln R$, and $F \equiv \mathcal{A}(\tau)(\ln R)^{2}$, and these coefficients are all non-negative.
By solving this equation with respect to $X(\tau)$, we obtain
\begin{align}
    X(\tau) \le \frac{E+\sqrt{E^2+4DF}}{2D} \le \frac{E}{D} + \frac{F}{E},
    \label{eq:Xtau_bound}
\end{align}
where we use $\sqrt{1+x}\le 1 + x / 2$ for $x\geq 0$ in the final inequality.
Substituting $X(\tau)=\Sigma(\tau) /\ln N-1$, $D$, $E$, and $F$ into this relation, we obtain Eq.~\eqref{eq:main_result_2} as follows:
\begin{align}
    \Sigma(\tau) &\le \mathcal{A}(\tau)\ln R\cdot\frac{R-1}{R+1} \nonumber\\
    &+\left(1+\frac{\ln R}{R+1} +\frac{\mathcal{A}(\tau)(\ln R)}{\mathcal{A}(\tau)(R-1) + \ln N }\right)\ln N \nonumber\\
    & \le  \mathcal{A}(\tau)\ln R\cdot\frac{R-1}{R+1}+\left(1+\frac{\ln R}{R+1} +\frac{\ln R}{R-1}\right)\ln N.
    \label{eq:non_ss_final_ineq}
\end{align}
This inequality satisfies the assumptions of Eq.~\eqref{eq:Sigma_assumption}.

\end{document}